\documentclass[pra,twocolumn,preprintnumbers,amsmath,amssymb,superscriptaddress,showpacs,longbibliography]{revtex4-1}
\usepackage{graphicx}
\usepackage{latexsym}
\usepackage{amsmath}
\usepackage{graphics}
\usepackage{amssymb}
\usepackage{layout}
\usepackage{verbatim}
\usepackage{amsfonts,epsfig,dsfont}
\usepackage{natbib}
\usepackage{hyperref}
\usepackage{soul}

\newcommand{\ket}[2][]{{|#2\rangle_{#1}}}
\newcommand{\bra}[2][]{{}_{#1}\langle #2|}

\newcommand{\Tr}{\textrm{Tr}}

\def\duzomniejsze{<\kern-.7mm<}
\def\duzowieksze{>\kern-.7mm>}

\def\textbf#1{{\bf #1}}
\def\beq{\begin{equation}}
\def\eeq{\end{equation}}
\def\be{\begin{equation}}
\def\ee{\end{equation}}
\def\ben{\begin{eqnarray}}
\def\een{\end{eqnarray}}
\def\beqa{\begin{eqnarray}}
\def\eeqa{\end{eqnarray}}
\def\eea{\end{array}}
\def\bea{\begin{array}}
	\newcommand{\unit}{\mathbb{I}}

\usepackage{color}
\definecolor{dgreen}{RGB}{0,90,0}

\begin{document}

\title{Qubit-environment entanglement generation and the spin echo}

\author{Katarzyna Roszak}
\affiliation{Department of Theoretical Physics, Faculty of Fundamental Problems of 
	Technology, Wroc{\l}aw University of Science and Technology,
50-370 Wroc{\l}aw, Poland}
\author{{\L}ukasz Cywi{\'n}ski}
\affiliation{Institute of Physics, Polish Academy of Sciences, 02-668 Warsaw, Poland}

\date{\today}

	\begin{abstract}
We analyze the relationship between qubit-environment entanglement that can be created during the pure dephasing of the qubit and the effectiveness of the spin echo protocol. We focus here on mixed states of the environment.   
We show that while the echo protocol can obviously counteract classical environmental noise, it can also undo dephasing associated with qubit-environment entanglement, and there is no obvious difference in its efficiency in these two cases. 
Additionally, we show that qubit-environment entanglement can be generated at the end of the echo protocol even when it is absent at the time of application of the local operation on the qubit (the $\pi$ pulse). We prove that this can occur only at isolated points in time, after fine-tuning of the echo protocol duration. Finally, we discuss the conditions under which the observation of specific features of the echo signal can serve as a witness of the entangling nature of the joint qubit-environment evolution.
	\end{abstract}

	\maketitle
	
	\section{Introduction \label{sec1}}
	Environmentally induced dephasing of superpositions of pointer states of a controlled quantum system	is commonly associated with creation of system-environment entanglement, or at least the presence of the latter is deemed to be necessary in order to call this process quantum decoherence \cite{Schlosshauer_book,Hornberger,Zurek_RMP03}. However, as has been pointed out in literature, this association holds only when the initial states of both the qubit {\it and the environment} are pure \cite{Kuebler_AP73,Schlosshauer_book,Zurek_RMP03,Hornberger}. In the more general, and much more realistic, case of mixed environmental states, dephasing of the system {\it does not} have to be accompanied by establishment of system-environment entanglement, and intuitions concerning distinguishing between ``quantum decoherence'' and ``dephasing due to classical environmental noise'' (understood here strictly as leading to no system-environment entanglement) that are built in works focusing on pure-state vs ``classical'' environments become unreliable \cite{Eisert_PRL02,Hilt_PRA09,Maziero_PRA10,Pernice_PRA11,Roszak_PRA15,Roszak_qudit_PRA18,Roszak_PRA18,Roszak_PRA19,Szankowski_arXiv20}.
	
We shed light on this general problem by focusing on the relationship between the effectiveness of qubit coherence recovery in a spin echo experiment \cite{Hahn_PR50,Abragam,Vandersypen_RMP05}, which is well known to lead to such a recovery when the environment is a source of external noise of mostly low-frequency character \cite{deSousa_TAP09,Szankowski_JPCM17}. We show that the echo procedure can (but does not have to) lead to coherence recovery when the dephasing is {\it not} associated with qubit-environment entanglement (QEE), but it can also undo QEE, while using only local operations on the qubit. Interestingly, there is no obvious correlation between the efficiency of coherence recovery and presence or absence of QEE generated during the evolution of the qubit and its environment.
 
In fact, we show that it is possible for QEE to appear at the end of the echo protocol, with no entanglement present at the time of application of the unitary operation to the qubit. 
This should not be surprising, as the evolutions that are most interesting in the context of echo protocol typically have non-Markovian character, and at the time of application of the local unitary operation the state of the qubit and the environment is typically correlated. 
	This effect can however only occur at isolated points in time, and this is the only feature of the echo experiment that conforms to the commonly encountered (but generally incorrect) intuitions that echo protocol should undo the generation of QEE, as it typically undoes qubit dephasing. 
	
	While most of our results underline the lack of strong correlation between efficacy of coherence recovery in the spin echo protocol and the presence of QEE during the evolution, we show that there is at least one situation in which the appearance of a phase shift between the initial and the echoed coherence of the qubit signifies that the evolution is of QEE-generating character.

	The paper is organized as follows. In Sec.~\ref{sec:PD_echo} we introduce the echo protocol for the qubit undergoing pure dephasing due to an interaction with its environment and recapitulate the basic criterion for appearance of QEE during pure dephasing evolution. In Sec.~\ref{sec:perfect} we discuss the conditions for the echo to work prefectly, i.e.~to lead to the recovery of the initial pure state of the qubit. As the perfect echo necessarily leads to the removal of any entanglement (if any was in fact present during the evolution),  in Section \ref{sec:imperfect_QEE} we focus on the imperfect echo and its relation to generation of entanglement during the evolution. There is no simple relation, and we show there that the echo can in fact lead to creation of entanglement in the final state even if there was none at the time of application of the local operation to the qubit. However, as we show in Section \ref{sec:isolated}, it can happen only at certain points in time, and the $\pi$ pulse applied to the qubit cannot transform a joint system evolution which is essentially nonentangling into an entangling one. Finally, in Sec.~\ref{sec:witness} we describe the conditions for the initial environmental state and qubit-environment coupling that allows to use the echo signal as a witness of the entangling nature of the evolution of the qubit and its environment. Sec.~\ref{sec5} concludes the paper.

	\section{Pure dephasing, entanglement, and echo} \label{sec:PD_echo}

	\subsection{Pure dephasing}
	In the following, we study the spin echo performed on a qubit in an arbitrary pure-dephasing scenario, meaning that the 
	only constraint on the 
	qubit-environment interaction is that it does not disturb the occupations of the qubit \cite{Roszak_PRA15,chen18,chen19}. The most general form of the Hamiltonian which describes the pure dephasing case is
	\begin{eqnarray}
	\label{ham}
	\hat{H}&=&
	\hat{H}_{\mathrm{Q}}+\hat{H}_{\mathrm{E}}+ |0\rangle\langle 0|\otimes{\hat{V}_0} +|1\rangle\langle 1|\otimes{\hat{V}_1} \label{eq:Hgen} \,\, .
	\end{eqnarray}
	The first term of the Hamiltonian describes the qubit and is given by $\hat{H}_{\mathrm{Q}}=\sum_{i=0,1}\varepsilon_{i}|i\rangle\langle i|$,  the second  describes the environment, while the remaining terms describe the qubit-environment interaction with the qubit states written on the left side of each term (the environment operators $\hat{V}_0$ and $\hat{V}_1$ are arbitrary, as is the free Hamiltonian of the environment $\hat{H}_{\mathrm{E}}$).
	Hence, the only constraint on the Hamiltonian, which restricts the qubit evolution
	to pure dephasing, is that the interaction term is diagonal with respect
	to the qubit eigenstates.
	
	The evolution operator corresponding to the Hamiltonian (\ref{ham})
	may in general be written in the form
	\begin{equation}
	\label{u}
	\hat{U}(t) = |0\rangle\langle 0|\otimes\hat{w}_0(t)+ |1\rangle\langle 1|\otimes \hat{w}_1(t) \,\, , 
	\end{equation}
	where $\hat{w}_i(t)= \exp(-\frac{i}{\hbar}\varepsilon_it)\exp(-\frac{i}{\hbar}\hat{H}_{i}t) $,
	with $\hat{H}_{i} \! =\! \hat{H}_{\mathrm{E}} + \hat{V}_{i}$ (the first exponential term is responsible for the phase evolution which comes from the free Hamiltonian of the qubit).
	Note that while $\hat{H}_{\mathrm{Q}}$ commutes with all the other terms in $\hat{H}$, this is not necessarily the case with $\hat{H}_{\mathrm{E}}$.
	We assume that the intial state has no correlations between the qubit and the environment, 
	\begin{equation}
	\label{ini}
	\hat{\sigma}(0)=|\psi\rangle\langle\psi |\otimes\hat{R}(0),
	\end{equation}
	with the initial qubit state $|\psi\rangle =a|0\rangle+b|1\rangle$
	and $\hat{R}(0)$ being the initial state of the environment. The qubit-environment density matrix at a later time can be written as
	\begin{equation}
	\label{mac0}
	\hat{\sigma}(t)=
	\left(
	\begin{array}{cc}
	|a|^2\hat{w}_0(t)\hat{R}(0)\hat{w}_0^{\dagger}(t)&ab^*
	\hat{w}_0(t)\hat{R}(0)\hat{w}_1^{\dagger}(t)\\
	a^*b
	\hat{w}_1(t)\hat{R}(0)\hat{w}_0^{\dagger}(t)
	&|b|^2\hat{w}_1(t)\hat{R}(0)\hat{w}_1^{\dagger}(t)
	\end{array}
	\right).
	\end{equation}
	Here the matrix form only pertains to the qubit subspace and is written in terms of qubit
	pointer states.
	If only the state of the qubit is of interest, then the reduced density matrix of the
	qubit is obtained by tracing out the environment from the matrix (\ref{mac0})
	and we get
	\begin{equation}
	\label{redmac0}
	\hat{\rho}(t)=\Tr_E\hat{\sigma}(t)=
	\left(
	\begin{array}{cc}
	|a|^2 &ab^*W(t)\\
a^*bW^*(t)
&|b|^2
\end{array}
\right),
\end{equation}
with normalized coherence 
\begin{equation}
\label{wu0}
W(t)=\Tr\left[ \hat{R}(0)\hat{w}_1^{\dagger}(t)\hat{w}_0(t)\right] \,\, .
\end{equation}

	\subsection{Spin echo during pure dephasing}
	
	The procedure which is known as the spin echo \cite{Hahn_PR50,Abragam,Vandersypen_RMP05} can be described 
	as follows. After the initialization of the qubit state, the qubit and
	environment evolve for a certain time $\tau$, after which a $\pi$-pulse about $x$ or $y$ axis is applied to the qubit (for concreteness we focus here on pulses about $x$ axis). Such a pulse interchanges the amplitudes of $\ket{0}$ and $\ket{1}$ states. Then the system is allowed to evolve for the same
	time period $\tau$ and another $\pi$-pulse is applied. In the ideal
	case, this leads to the qubit regaining its initial state at time 
	$2\tau$ (after the second $\pi$-pulse), but even in non-ideal scenarios
	the decoherence which is observed after the echo sequence 
	can be much smaller compared to the evolution without the echo when the environment is a source of external noise of mostly low-frequency character \cite{deSousa_TAP09} (see Section \ref{sec:small} below for a concise formal explanation of this fact).
	
	The evolution in the echo experiment with the final time $2\tau$ is described by the operator
	\begin{equation}
	\label{echo}
	\hat{U}_{\mathrm{echo}}(2\tau)=
	\hat{\sigma}_x \hat{U}(\tau)\hat{\sigma}_x \hat{U}(\tau),
	\end{equation}
	where $\hat{\sigma}_x$ is the appropriate Pauli matrix which describes 
	the action of the $\pi$-pulse on the qubit
	and $\hat{U}(\tau)$ is a joint system-environment evolution operator,
	which for pure dephasing is given by eq.~(\ref{u}). The second $\pi$ pulse at time $2\tau$ interchanges the two complex-conjugate coherences in the final reduced state of the qubit.
	and it is added for convenience, to make the final coherence equal to the original one, not to its complex conjugate, in the case of perfect echo. 
	
	We assume that 
	the initial state of the qubit-environment system 
	is given by eq.~(\ref{ini}).
	Then the joint system-environment state at time $\tau$ before the first $\pi$-pulse
	is given by the desity matrix (\ref{mac0}).
	Modeling the whole procedure with the evolution operator (\ref{echo})
	we get the qubit-environment state directly after the 
	echo sequence is performed, which is given by
\begin{widetext}
\begin{equation}
\label{mac}
\hat{\sigma}(2\tau)=
\left(
\begin{array}{cc}
|a|^2\hat{w}_1(\tau)\hat{w}_0(\tau)\hat{R}(0)\hat{w}_0^{\dagger}(\tau)\hat{w}_1^{\dagger}(\tau)&ab^*
\hat{w}_1(\tau)\hat{w}_0(\tau)\hat{R}(0)\hat{w}_1^{\dagger}(\tau)\hat{w}_0^{\dagger}(\tau)\\
a^*b
\hat{w}_0(\tau)\hat{w}_1(\tau)\hat{R}(0)\hat{w}_0^{\dagger}(\tau)\hat{w}_1^{\dagger}(\tau)
&|b|^2\hat{w}_0(\tau)\hat{w}_1(\tau)\hat{R}(0)\hat{w}_1^{\dagger}(\tau)\hat{w}_0^{\dagger}(\tau)
\end{array}
\right).
\end{equation}
\end{widetext}
The echoed qubit state is obtained,
as in the case of simple decoherence (\ref{redmac0}), by tracing out the environment from eq.~(\ref{mac}),
which yields $\hat{\rho}(2\tau)=\Tr_E\hat{\sigma}(2\tau)$, which has the same structure as eq.~(\ref{redmac0}), but with normalized coherence
\begin{equation}
\label{wu}
W(2\tau)=\Tr\left[\hat{R}(0) \hat{w}_1^{\dagger}(\tau)\hat{w}_0^{\dagger}(\tau)
\hat{w}_1(\tau)\hat{w}_0(\tau)\right] \,\, .
\end{equation}

\subsection{QEE condition for pure dephasing with and without echo}
For any bipartite density matrix which can be written in the form (\ref{mac0}), the if and only if
condition of qubit-environment separability is
\begin{equation}
\label{war_ent}
[\hat{w}_0^{\dagger}(t)\hat{w}_1(t),\hat{R}(0)]=0 \,\, ,
\end{equation}
as has been proven in Ref.~\cite{Roszak_PRA15}.
The original derivation involves the positive-partial-transpose (PPT) criterion \cite{Peres_PRL96,Horodecki_PLA96}
in one direction and the definition of mixed bipartite separable states in the other.
Since the qubit-environment state at time $\tau$ before the $\pi$-pulse is applied is given precisely 	
	by eq.~(\ref{mac0}), the condition can be explicitly used to check for QEE present just before the application of the pulse (the pre-pulse entanglement).
	
	The QEE present in the system after the echo procedure
	is performed is similarly straightforward to study, because the qubit-environment
	density matrix (\ref{mac}) is of the same form as the one that is obtained by a simple pure-dephasing interaction (\ref{mac0}).
	The two can be reduced to one another by the transformation
	\begin{subequations}
		\begin{eqnarray}
		\hat{w}_0'(2\tau)&=&\hat{w}_1(\tau)\hat{w}_0(\tau),\\
		\hat{w}_1'(2\tau)&=&\hat{w}_0(\tau)\hat{w}_1(\tau).
		\end{eqnarray}
	\end{subequations}
	Then the condition for separability of the 
	echoed state is
	\begin{equation}
	\label{war_ent2}
	[\hat{w}_0'^{\dagger}(2\tau)\hat{w}_1'(2\tau),\hat{R}(0)]=
	[\hat{w}_0^{\dagger}(\tau)\hat{w}_1^{\dagger}(\tau)\hat{w}_0(\tau)\hat{w}_1(\tau),\hat{R}(0)]=0.
	\end{equation}

	\section{Conditions for perfect echo}  \label{sec:perfect}
	\subsection{General considerations}
For the echo to be perfect, meaning that the qubit state which is obtained after performing the echo  is equal to the initial qubit state, $\Tr_E \hat{\sigma}(2\tau)=|\psi\rangle\langle\psi |$, the following condition needs to be met,
	\begin{equation}
		\label{war_perfect}
		[\hat{w}_0^{\dagger}(\tau),\hat{w}_1(\tau)]=0.
	\end{equation}
	The complementary condition 
	$[\hat{w}_0(\tau),\hat{w}_1(\tau)]=0$ follows from the above equation,
	since commutation of two operators implies that there exists a basis in which 
	both operators are diagonal and the Hermitian conjugate of any operator
	is always diagonal in the same basis as the operator itself.
	
	In the situation when the echo reinstates the initial qubit state,
	it also severs any entanglement which may have been generated between
	the qubit and the environment during their joint evolution.
	However, the condition for perfect echo is not related in any way to the condition
	for absence of QEE at time $\tau$, which is given by eq.~(\ref{war_ent}). The latter depends on the initial state of the density matrix of the environment and can be fulfilled both when
	the conditional evolution operators of the environment commute, and when they do not.
	
	It is fairly straightforward to find an evolution which leads to a perfect echo for a given $\tau$, or even for any $\tau$, but does not
	lead to any QEE generation,
	and one that does lead to entanglement generation. 
For example, if $[\hat{V}_{i},\hat{H}_{E}]\! =\! 0$ for $i\!= \! 0$, $1$, and $\hat{R}(0)\propto \exp(-\beta\hat{H}_{E})$, i.e.~the environment is in a thermal equlibrium state achieved in absence of the qubit, then there is no entanglement generated at time $\tau$, as eq.~(\ref{war_ent}) is fulfilled. However, the echo is perfect only if additionally $[\hat{V}_0,\hat{V}_1]\! = \! 0$. 

On the other hand, if we assume all the commutation relations from the previous example to be fulfilled, but take $\hat{R}(0)$ such that $[\hat{R}(0),\hat{V}_0-\hat{V}_1] \! \neq \! 0$, we have perfect echo at time $2\tau$, but the qubit-environment state is entangled at time $\tau$. These examples already show that the behavior of ``echoed'' coherence reflects the general feature of dephasing caused by an environment in a mixed state: there is no direct correspondence between the generation of QEE and the amount of dephasing. The echo procedure can undo dephasing (even perfectly) not only in the ``classical dephasing'' case (using the terminology from Ref.~\cite{Schlosshauer_book}), in which no entanglement is established, but also in the ``true quantum decoherence'' case, in which entanglement is created during the evolution.

	
\subsection{Small decoherence limit}  \label{sec:small}
If the echoed coherence $W(2\tau)$ is close to unity, as happens when $2\tau$ is close to the time at which the echo is perfect, one can approximate it by an expression valid to second-order in qubit-environment coupling. For simplicity, let us focus on a less general form of the $\hat{V}_i$ operators, namely
\begin{align}
\hat{V}_0 &= \frac{1}{2} \lambda (\eta +1)\hat{V} \,\,   ,\nonumber\\
\hat{V}_1 &= \frac{1}{2} \lambda (\eta -1)\hat{V} \,\, ,
\end{align}
so that the qubit-environment coupling takes the form $\frac{1}{2}\lambda(\eta\hat{\mathds{1}} - \hat\sigma_z)\otimes\hat V$. In the formulas above, $\lambda$ is a dimensionless parameter controlling the strength of the coupling, while $\eta$ controls the ``bias'' of the coupling. A commonly used ``unbiased'' coupling, $\propto \hat\sigma_z\otimes\hat V$, which occurs for example for qubits based on spin-$1/2$ entities coupled to an environment via the magnetic dipole interaction \cite{Cywinski_APPA11,Chekhovich_NM13}, corresponds to  $\eta\!=\! 0$, while the ``biased'' case of $\eta\! = \! -1$ applies for example to excitonic qubits \cite{borri01,vagov03,vagov04,roszak06b}, or to qubits based on $m\! =\! 0$ and $m\! =\! \pm 1$ levels of a qubit based on a spin-$1$ entity such as a nitrogen-vacancy center in diamond \cite{Zhao_PRB12,Kwiatkowski_PRB18}. A calculation of coherence up to $\lambda^2$ order gives \cite{Paz_PRA17,Kwiatkowski_PRB20}
\begin{equation}
W(2\tau) \approx 1 -\lambda^2 \chi(2\tau) - i\eta\lambda^2\Phi(2\tau) \,\, ,
\end{equation}
where the attenuation function $\chi(t)$ and the phase shift $\Phi(t)$ are real functions given by  
\begin{align}
\chi(2\tau) & =  \frac{1}{2} \int_{0}^{2\tau} \mathrm{d}t_1 \int_{0}^{t_1}\mathrm{d}t_2 f(t_1) f(t_2)  C(t_1,t_2) \,\, , \label{eq:chit}\\
\Phi(2\tau) & =  \frac{1}{2} \int_{0}^{2\tau} \mathrm{d}t_1 \int_{0}^{t_1}\mathrm{d}t_2 f(t_2) K(t_1,t_2) \,\, , \label{eq:Phit} 
\end{align}
where 
\begin{equation}
C(t_1,t_2)  =  \mathrm{Tr}_{E}\left( \hat{R}(0) \{ \hat{V}(t_1),\hat{V}(t_2)\} \right) \,\, 
\end{equation}
is the autocorrelation function of the operator $\hat{V}(t) = \exp(i\hat{H}_{E}t)\hat{V}\exp(-i\hat{H}_{E}t)$, while
\begin{equation}
K(t_1,t_2)  = -i\theta(t_1-t_2) \mathrm{Tr}_{E}\left( \hat{R}(0) [ \hat{V}(t_1),\hat{V}(t_2)] \right) \,\, 
\end{equation}
 is the linear response function \cite{Negele,Bruus} associated with this operator, and the temporal filter function \cite{deSousa_TAP09,Cywinski_PRB08} for the echo experiment is given by $f(t) = \Theta(t)\Theta(\tau-t) - \Theta(t-\tau)\Theta(2\tau-t)$, i.e.~$|f(t)|\!=\!1$ for $t\! \in\! [0,2\tau]$, is zero otherwise, and changes sign at $t\! =\! \tau$.  For the derivation of the expression for $\chi(2\tau)$ see Ref.~\cite{Szankowski_JPCM17}, while the derivations of the formula for phase $\Phi(2\tau)$ can be found in Refs \cite{Paz_PRA17} and \cite{Kwiatkowski_PRB20}.

We assume that the environment is initially in a stationary state of its free Hamiltonian, $[\hat{R}(0),\hat{H}_{E}]\! =\! 0$, which implies that $C(t_1,t_2)$ is a function of a single variable, $\Delta t = t_1 - t_2$. We can then introduce the power spectral density (PSD) of the noise, defined by
\beq
S(\omega) = \int_{-\infty}^{\infty} e^{i\omega \Delta t} C(\Delta t) \mathrm{d}\Delta t \,\, , \label{eq:S}
\eeq
and express the attenuation function and the phase shift as 
\begin{align}
\chi(2\tau) & =  \int_{-\infty}^{\infty} \frac{8 \sin^4 \frac{\omega \tau}{2}}{\omega^2}  S(\omega) \frac{\mathrm{d}\omega}{2\pi} \,\, , \label{eq:chiw}\\
\Phi(2\tau) & = \int_{-\infty}^{\infty} \frac{8 \sin^4 \frac{\omega \tau}{2}}{\omega^2}  \mathrm{cotan} \frac{\omega \tau}{2} \mathrm{tanh} \frac{\beta \omega}{2} S(\omega) \frac{\mathrm{d}\omega}{2\pi}  \,\, . \label{eq:Phiw} 
\end{align}
Here, in order to derive the second of these expressions, we have assumed that the environment is actually in a thermal state, i.e.~$\hat{R}(0) \! =\! e^{-\beta\hat{H}_{E}}/\mathrm{Tr}e^{-\beta\hat{H}_{E}}$. 

Vanishing $\chi(2\tau)$ is necessary for the perfect echo. 
	Here we see that, taking into account the fact that $S(\omega)$ is positive-definite, this can happen at $\tau \! \neq \! 0$ only when PSD consists of a series of delta peaks at frequencies $\omega_k \! =\! 2\pi k/\tau$ for integer $k$. The most commonly encountered case is of PSD concentrated only at very low frequencies (only the $k\!=\! 0$ peak is present), i.e.~$S(\omega) \! \propto \!  \delta(\omega)$. This corresponds to a time-independent symmetric correlator of $\hat{V}(t)$, i.e.~$C(\Delta t)$, which requires $[\hat{H}_E,\hat{V}]\! =\! 0$. This situation is thus equivalent to the previously discussed case of the perfect echo. The situation of $S(\omega)$ with periodically positioned narrow peaks in frequency is more interesting, as it corresponds to $\hat{V}(t)$ that has nontrivial dynamics. It is also not particularly artificial: it corresponds to situations in which the second-order correlation function of the environmental operator $\hat{V}$ has a well-defined periodicity. A perfect echo can occur at isolated points in time in this case.

Let us note that while the response function $K(\Delta t)$ vanishes when the environment is completely mixed, the symmetric correlation function $C(\Delta t)$ has no reason to vanish in this situation. The presence of a finite attenuation function $\chi$, and thus of finite decay of qubit coherence, obviously does not require the presence of QEE: note that the condition (\ref{war_ent}) for Q-E separability is fulfilled for a completely mixed initial environmental state.


	\section{Imperfect echo and QEE \label{sec:imperfect_QEE}}
	
	\subsection{Echo-induced entanglement \label{sec}}

	Let us consider the situation when at time $\tau$, at which we apply a {\it local} operation to one part (the qubit) of our bipartite system,  the condition 
	of qubit-environment separability is fulfilled (\ref{war_ent}), 
	but the perfect-echo condition (\ref{war_perfect}) is not.
		Since the perfect echo kills any QEE that was generated during the evolution, one could expect that a non-perfect echo, still leading to a partial recovery of coherence, should diminish its amount compared to values attained during the evolution, for example at the time of application of the pulse. 
	In particular, if the evolution does not entangle the qubit with
	is environment at the time the first $\pi$-pulse is applied, it should not lead to QEE 
	after the whole echo procedure is performed.
	In the following, we will show 	that this is in fact not necessarily the case. 
	This is nothing else, but another result of the general fact that the magnitude of system dephasing is rather weakly affected by presence or absence of system-environment entanglement {\it when the environmental state is far from being pure}. 
	
	The condition of separability (\ref{war_ent})
	is equivalent to the statement that there exists a basis in which both the
	operator $\hat{w}_0^{\dagger}(\tau)\hat{w}_1(\tau)$ and the initial density
	matrix of the environment $\hat{R}(0)$ are diagonal.
	Although diagonality in this basis is obviously preserved for the conjugate
	of $\hat{w}_0^{\dagger}(\tau)\hat{w}_1(\tau)$ 
	there is no reason why the operators
	$\hat{w}_1^{\dagger}(\tau)$ and $\hat{w}_0(\tau)$ should also be diagonal 
	in this basis. 
	In other words, for any two evolution operators $\hat{w}_0^{\dagger}(\tau)$ and 
	$\hat{w}_1(\tau)$ which do not commute at a given time $\tau$
	(which means that $\hat{w}_0^{\dagger}(\tau)$ is diagonal in a 
	different basis than
	$\hat{w}_1(\tau)$),
	there exists a set of initial environmental states for which
	$[\hat{w}_0^{\dagger}(\tau)\hat{w}_1(\tau),\hat{R}(0)]=0$.
	If the initial state of the environment is described by one of these density
	matrices then
	at time $\tau$ (both before and after the first $\pi$-pulse), the 
	qubit-environment density matrix obtained by using the evolution operator
	(\ref{u}) is separable, but is no longer a product state.
	The state (after the $\pi$-pulse) can be written as
	\begin{equation}
	\label{wechu}
	\sigma(\tau)=\left(
	\begin{array}{cc}
	|b|^2\hat{R}_{00}(\tau)&
	a^*b\hat{w}_1(\tau)\hat{w}_0^{\dagger}(\tau)\hat{R}_{00}(\tau)
	\\
	ab^*\hat{R}_{00}(\tau)\hat{w}_0(\tau)\hat{w}_1^{\dagger}(\tau)
	&|a|^2\hat{R}_{00}(\tau)
	\end{array}
	\right),
	\end{equation}
	where 
	$\hat{R}_{00}(\tau)=\hat{w}_0(\tau)\hat{R}(0)\hat{w}_0^{\dagger}(\tau)$
	and the fact that 
	\begin{equation}
	\label{wrw}
	\hat{w}_0(\tau)\hat{R}(0)\hat{w}_0^{\dagger}(\tau)=
	\hat{w}_1(\tau)\hat{R}(0)\hat{w}_1^{\dagger}(\tau)
	\end{equation}
	is a straightforward consequence of the separability criterion (\ref{war_ent})
	being fulfilled at time $\tau$.
	Applying the other half of the echo procedure (unitary evolution
	$U(\tau)$ followed by the $\sigma_x$ operator) yields
	\begin{widetext}
		\begin{equation}
		\label{poechu}
		\sigma(2\tau)=\left(
		\begin{array}{cc}
		|a|^2\hat{w}_1(\tau)\hat{R}_{00}(\tau)\hat{w}_1^{\dagger}(\tau)&
		ab^*\hat{w}_1(\tau)\hat{R}_{00}(\tau)\hat{w}_0(\tau)\hat{w}_1^{\dagger}(\tau)
		\hat{w}_0^{\dagger}(\tau)
		\\
		a^*b\hat{w}_0(\tau)\hat{w}_1(\tau)\hat{w}_0^{\dagger}(\tau)\hat{R}_{00}(\tau)
		\hat{w}_1^{\dagger}(\tau)
		&|b|^2\hat{w}_0(\tau)\hat{R}_{00}(\tau)\hat{w}_0^{\dagger}(\tau)
		\end{array}
		\right).
		\end{equation}
	\end{widetext}
	This qubit-environment density matrix is separable, if and only if
	the condition
	\begin{equation}
	\label{cond_echo}
	\left[\hat{w}_0^{\dagger}(\tau)\hat{w}_1(\tau),\hat{R}_{00}(\tau)\right]=0
	\end{equation}
	is fulfilled.
	The condition is equivalent
	to the separability criterion for a product initial state of the qubit and
	the environment initially in state $\hat{R}_{00}(\tau)$, when the evolution is governed by the operators
	$\hat{w}_0(\tau)$ and $\hat{w}_1(\tau)$, eq.~(\ref{war_ent}).
	Interestingly,	
	the resulting state (\ref{poechu}) is different than the state which would be obtained at time $\tau$
	from an initial environmental state $\hat{R}(0)=\hat{R}_{00}(\tau)$.
	This becomes obvious
	when the elements of the density matrix proportional to $ab^*$ are 
	compared in both cases, since $\hat{w}_0(\tau)\hat{w}_1^{\dagger}(\tau)
	\hat{w}_0^{\dagger}(\tau) \neq \hat{w}_1^{\dagger}(\tau)$
	(because we assumed that $\hat{w}_0(\tau)$ and $\hat{w}_0^{\dagger}(\tau)$
	do not commute with $\hat{w}_1^{\dagger}(\tau)$).

	\subsection{Example of qubit-environment
		entanglement generated via
		the spin echo at time $2\tau$ for separable state at time $\tau$
		\label{ex}}
	
	As an example let us study a qubit interacting with an environment of dimension
	$N=2$. We will study a pair of interaction operators $\hat{w}_0(\tau)$
	and $\hat{w}_1(\tau)$ that do not lead to entanglement in the density matrix
	(\ref{wechu}), but lead to entanglement in the echoed density matrix
	(\ref{poechu}) for a set of initial environmental states.
	
	Our exemplary operators $\hat{w}_0(\tau)$ and $\hat{w}_1(\tau)$
	written in the eigenbasis of the initial environment density matrix
	$\hat{R}(0)=c_{0} \ket{0}\bra{0}+c_{1} \ket{1}\bra{1}$
	are
	\begin{subequations}
		\begin{eqnarray}
		\label{w0+}
		\hat{w}_0^{\dagger}(\tau)&=&\frac{1}{\sqrt{2}}\left(
		\begin{array}{cc}
		1&1\\
		-1&1
		\end{array}\right),\\
		\label{w1}
		\hat{w}_1(\tau)&=&\hat{w}_1^{\dagger}(\tau)=\frac{1}{\sqrt{2}}\left(
		\begin{array}{cc}
		1&1\\
		1&-1
		\end{array}\right).
		\end{eqnarray}
	\end{subequations}
	The operators do not commute and we find that
	\begin{equation}
	\hat{w}_0^{\dagger}(\tau)\hat{w}_1(\tau)
	=\hat{w}_1^{\dagger}(\tau)\hat{w}_0(\tau)
	=\left(
	\begin{array}{cc}
	1&0\\
	0&-1
	\end{array}\right)
	\end{equation}
	are diagonal in the eigenbasis of $\hat{R}(0)$
	meaning that the evolution (without the echo)
	does not yield entanglement at time $\tau$
	for any $c_0$,
	since $[\hat{w}_0^{\dagger}(\tau)\hat{w}_1(\tau),\hat{R}(0)]=0$.
	On the other hand, this does not mean that there is no qubit decoherence,
	since the off-diagonal elements of the qubit density matrix are proportional to
	\begin{equation}
	\mathrm{Tr}\left[\hat{w}_1^{\dagger}(\tau)\hat{w}_0(\tau)\hat{R}(0)\right]=
	c_0-c_1.
	\end{equation}
	Hence, the qubit state remains pure only for an initial pure state of the environment,
	$c_0=0$ or $1$, with the purity reaching its minimal possible value in the 
	type of evolutions described for a completely mixed environment, $c_0=c_1=1/2$.
	
	It is now straightforward to find the operators 
	which govern QEE in the case of the quantum
	echo,
	\begin{equation}
	\hat{w}_0^{\dagger}(\tau)\hat{w}_1^{\dagger}(\tau)\hat{w}_0(\tau)\hat{w}_1(\tau)
	=\left(
	\begin{array}{cc}
	0&1\\
	-1&0
	\end{array}\right).
	\end{equation}
	This operator is obviously not diagonal in the eigenbasis of the initial environment density matrix. Furthermore,
	\begin{equation}
	\left[\hat{w}_0^{\dagger}(\tau)\hat{w}_1^{\dagger}(\tau)\hat{w}_0(\tau)\hat{w}_1(\tau),\hat{R}(0)\right]
	=(c_1-c_0)\left(
	\begin{array}{cc}
	0&1\\
	1&0
	\end{array}\right)
	\end{equation}
	and the condition for separability (\ref{war_ent2})
	is fulfilled only for $c_0=c_1=\frac{1}{2}$, another words,
	only when the initial density matrix of the environment is proportional
	to unity, $\hat{R}(0)\sim\unit$. 
	
	When it comes to qubit decoherence, we always have
	\begin{equation}
	\mathrm{Tr}\left[\hat{w}_1^{\dagger}(\tau)\hat{w}_0^{\dagger}(\tau)\hat{w}_1(\tau)\hat{w}_0(\tau)\hat{R}(0)\right]=0,
	\end{equation}
	which means that the qubit at time $2\tau$ is always fully decohered, regardless 
	of the initial state of the environment. In this extreme case, the spin echo can do no damage in the best scenario, while
	for most states of the environments, the procedure strongly enhances decoherence. This should not be surprising in light of discussion from Sec.~\ref{sec:small}, as for such a small (two-dimensional) environment the correlation function of any environmental operator has to be periodic. 
	
	This example shows that 
	the echo may lead to the increase of entanglement
	with respect to the entanglement present in the system at the end of the
	free-evolution period in the echo procedure (since it can create such entanglement).
	This is contrary to intuition, since it is natural to try to extend
	the notion, that since a perfect echo procedure diminishes all
	QEE (while diminishing all decoherence),
	an imperfect echo should lead to lesser entanglement while it leads to lesser
	decoherence in the echoed state. As we see here, there exist situations
	when the echo not only increases entanglement, but also increases decoherence,
	and can be counterproductive. 	
	Using the physical picture discussed for weak dephasing in Sec.~\ref{sec:small} (and taking it strictly speaking outside of domain of its quantitative applicability, unless we assume a Gaussian environment \cite{Szankowski_JPCM17} for which $|W(2\tau)| \! =\! \exp[-\chi(2\tau)]$), we see that this can occur when the PSD of the environmental noise is periodic, but $\tau$ is such that it is the {\it maximum} of the filter $|\tilde{f}(\omega)|^2$ in eq.~(\ref{eq:chiw}) that overlaps with the peaks of $S(\omega)$.

	\subsection{Entangling evolution - pure environmental states}
	
	Let us study the special
	case of a pure initial state of the environment (we expect from the results of 
	the previous subsection that this situation will enhance the differences
	between the pre-pulse entanglement and echoed entanglement). 
	Then the joint state of the
	system and the environment is pure at any time, so it is pure at time 
	$\tau$ (pre-pulse) and at echo time $2\tau$. In this situation, entanglement at any time can be evaluated
	in a straightforward manner using the von Neumann entropy of one of the entangled
	subsystems, which is a good entanglement measure for pure states.
	The measure is defined as
	\begin{equation}
	\label{von}
	E(|\psi(t)\rangle)=-\frac{1}{\ln 2}\Tr\left(
	\rho(t)\ln\rho(t)
	\right),
	\end{equation}
	where $|\psi(t)\rangle$ is the pure system-environment state
	so $\sigma(t)=|\psi(t)\rangle\langle \psi(t) |$, $\rho(t)=\Tr_E|\psi(t)\rangle\langle \psi(t)|$
	is the density matrix of the qubit at time $t$ (obtained by tracing out the
	environment), and the entanglement measure is 
	normalized to yield unity for maximally entangled states.
	The same result would be obtained when tracing out the qubit degrees of freedom
	instead of the environmental degrees of freedom, but the small dimensionality
	of the qubit makes this way much more convenient.
	
	Let us denote
	the pure initial state of the environment as $|R_0\rangle$. Then
	qubit-environment state at time $\tau$ (pre-pulse) is given by
	\begin{equation}
	|\psi(\tau)\rangle = a|0\rangle \otimes\hat{w}_0(\tau)|R_0\rangle+b|1\rangle \otimes\hat{w}_1(\tau)|R_0\rangle
	\end{equation}
	and the corresponding echoed state (at time $2\tau$) is
	\begin{equation}
	|\psi(2\tau)\rangle = a|0\rangle \otimes\hat{w}_1(\tau)\hat{w}_0(\tau)|R_0\rangle+b|1\rangle \otimes\hat{w}_0(\tau)\hat{w}_1(\tau)|R_0\rangle.
	\end{equation}
	The qubit density matrices are then of the general form (\ref{redmac0})
	with $W(\tau)=\langle R_0|\hat{w}_1^{\dagger}(\tau)\hat{w}_0(\tau)|R_0\rangle$
	pre-pulse, and $W(2\tau)=\langle R_0|\hat{w}_1^{\dagger}(\tau)\hat{w}_0^{\dagger}(\tau)\hat{w}_1(\tau)
	\hat{w}_0(\tau)|R_0\rangle$
	for the echoed state. Hence, the absolute values of functions $W(\tau)$ and $W(2\tau)$
	constitute the degrees of coherence retained in the qubit system at the time of application of the pulse and
	at the echo time, respectively.
	
	The entanglement measure of eq.~(\ref{von}) can be calculated using eq.~(\ref{redmac0})
	which yields
	\begin{eqnarray}
	E(|\psi(t)\rangle)&=&-\frac{1}{\ln 2}
	\left[
	\frac{1+\sqrt{\Delta(t)}}{2}\ln\frac{1+\sqrt{\Delta(t)}}{2}\right.\\
	\nonumber
	&&+\left.
	\frac{1-\sqrt{\Delta(t)}}{2}\ln\frac{1-\sqrt{\Delta(t)}}{2}
	\right],
	\end{eqnarray}
	with $\Delta(t)=1-4|a|^2|b|^2+|a|^2|b|^2|W(t)|^2$.
	Note that $\Delta(t)$ is an increasing function of the degree of coherence
	$|W(t)|$, while entanglement measured by $E(|\psi(t)\rangle)$
	is a decreasing function of $\Delta(t)$, so entanglement
	is a decreasing function of coherence $|W(t)|$, which means (as expected) that the 
	higher the qubit coherence, the lower the QEE.
	Consequently, the situation described at the beginning of Sec.~\ref{sec}, when the pre-pulse state
	$\sigma(\tau)$ has no QEE, but the echoed state
	$\sigma(2\tau)$ is entangled, for a pure initial state of the environment
	translates to the pre-pulse qubit state being more coherent than the echoed 
	qubit state, meaning that the echo can have an opposite effect on the qubit
	coherence than intended. This should be kept in mind when dealing with rather small environments that have a discrete spectrum, and which are close to being in pure state (e.g.~their temperature is very low, or, in case of spin environments, a large nonequilibrium polarization of the environmental spins was previously established, see Ref.~\cite{Roszak_PRA19} for discussion of QEE in this case).

	\begin{figure}[th]
		\begin{center}
			\unitlength 1mm
			\begin{picture}(75,55)(5,5)
			\put(0,0){\resizebox{85mm}{!}{\includegraphics{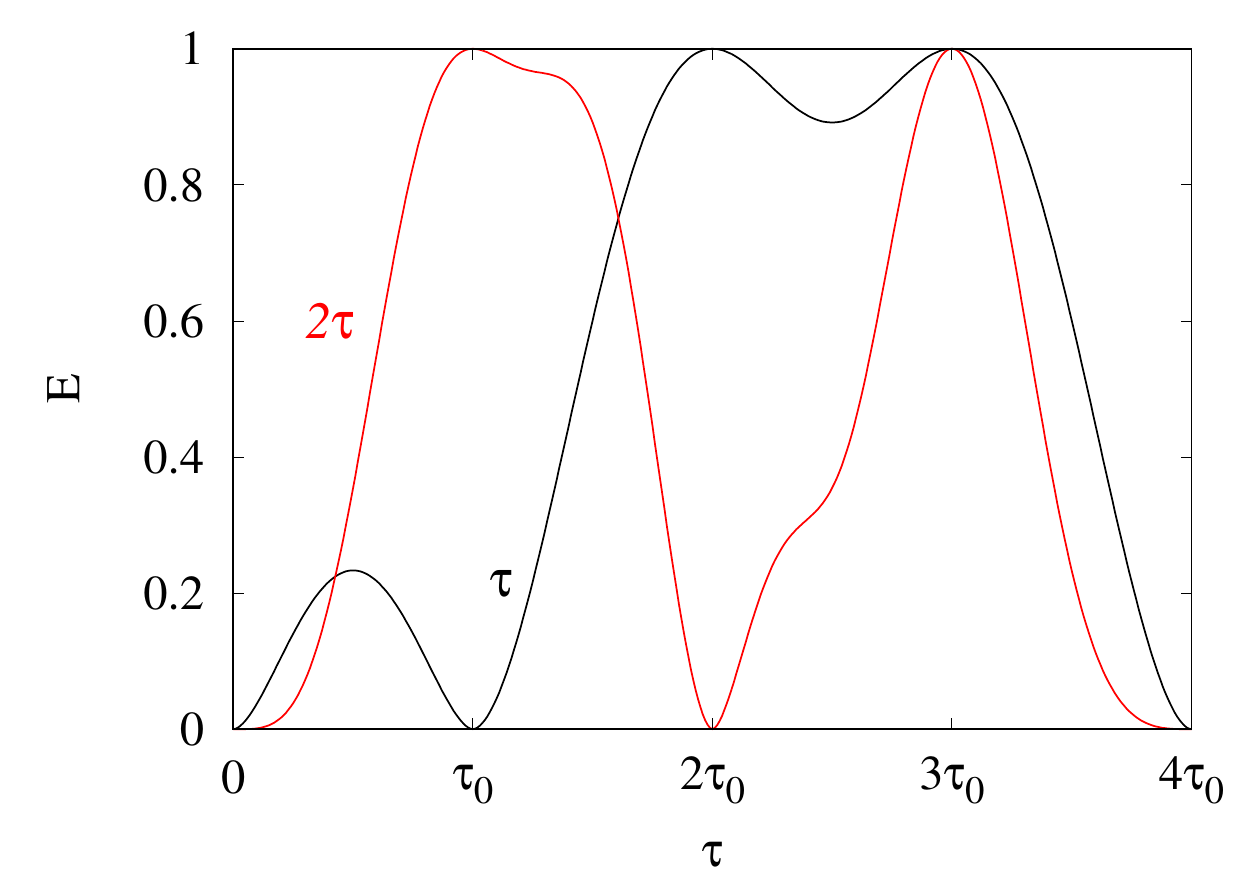}}}
			\end{picture}
		\end{center}
		\caption{\label{fig0} Exemplary QEE evolution 
			for a single qubit
			environment initially in a pure state pre-pulse (at time $\tau$, black
			line) and the corresponding echoed entanglement (at time $2\tau$,
			red line).}
	\end{figure}
	
	Fig.~(\ref{fig0}) shows an exemplary evolution of the QEE,
	measured by the normalized von Neumann entropy of eq.~(\ref{von}),
	for an environment restricted to a single qubit which is initially in a pure state.
	The evolution operators (in the subspace of the environment) are given by
	\begin{equation}
	\hat{w}_i(t)= e^{i\omega_i t}|\psi_i\rangle\langle\psi_i |
	+e^{i\omega_i' t}|\psi_i'\rangle\langle\psi_i' |,
	\end{equation}
	with $i=0,1$, $\omega_0=\pi/(4\tau_0)$, $\omega_0'=-\pi/(4\tau_0)$, $\omega_1=\pi/\tau_0$, $\omega_1'=2\pi/\tau_0$, and 
	\begin{eqnarray}
	|\psi_0\rangle&=&\frac{1}{\sqrt{2}}|R_0\rangle-\frac{i}{\sqrt{2}}|R_1\rangle,\\
	|\psi_0'\rangle&=&\frac{1}{\sqrt{2}}|R_0\rangle+\frac{i}{\sqrt{2}}|R_1\rangle,\\
	|\psi_1\rangle&=&\frac{\sqrt{2-\sqrt{2}}}{2}|R_1\rangle-
	\frac{\sqrt{2+\sqrt{2}}}{2}|R_0\rangle,\\
	|\psi_1'\rangle&=&\frac{\sqrt{2+\sqrt{2}}}{2}|R_0\rangle+
	\frac{\sqrt{2-\sqrt{2}}}{2}|R_1\rangle,
	\end{eqnarray}
	where $|R_1\rangle$ is the state perpendicular to the initial environmental state
	$|R_0\rangle$. Obviously, the evolution is periodic and repeats itself every 
	$4\tau_0$, while at $t=\tau_0$ the evolution operators are equal to the operators
	introduced in Sec.~\ref{ex}, for which a non-entangled state before the pulse
	leads to an entangled echoed qubit-environment state.
	
	The black line in Fig.~(\ref{fig0}) (denoted as $\tau$) shows the amount of entanglement between 
	the qubit and the environment as a function of time $\tau$, when no echo is performed.
	The red line (denoted as $2\tau$), on the other hand, shows qubit-environment
	entanglement at time $2\tau$ in the situation when a $\pi$ pulse was applied to the qubit
	at time $\tau$, again as a function of $\tau$. Hence, the two curves in Fig.~(\ref{fig0})
	show pre-pulse entanglement and the corresponding echoed entanglement
	as a function of the same parameter $\tau$.	
	The evolution of echoed entanglement is much more involved, and the interplay
	of the two curves shows that apart from the previously predicted $\tau=\tau_0$
	case (when no pre-pulse entanglement is observed, but there is echoed entanglement),
	there are many situations when applying the pulse enhances qubit-environment
	entanglement at a later time.
	Note, that for a pure initial state of the environment, 
	there is a strict correspondence between QEE and qubit
	coherence,
	meaning that every time entanglement is enhanced by the echo, the coherence of 
	the qubit is damped, and the effect of the echo is contrary to its purpose.
	
	\section{Echo induced entanglement is not possible for principally nonentangling evolutions} \label{sec:isolated}
	
	Although the examples discussed above show that the spin echo procedure can lead
	to the appearance of QEE at echo time when the qubit-environment state  was separable before the application of the pulse to the qubit,
	this occurs in rather special situations.
	Let us show now that it is only possible at isolated points of time, and
	there are no finite time intervals $t\in[\tau_1,\tau_2]$
	for which the pre-pulse state $\hat{\rho}(t)$ is separable, while the echoed state
	$\hat{\rho}(2t)$ is entangled. Since this is the case, we can extend the time interval
	to encompass the whole pre-pulse evolution $t\in[0,\infty]$, which yields the result
	that the echo procedure cannot be used to modify a non-entangling evolution
	into an entangling one.
	
	The argument is as follows. Separable evolutions, which obviously must fulfill
	the criterion (\ref{wrw}), can be divided into two categories:
	One encompasses all types of evolutions for which the environment does not 
	evolve, 
	\begin{equation}
	\Tr_Q\hat{\sigma}(t)=\hat{R}_{00}(t)=\hat{R}_{11}(t)=\hat{R}(0).
	\end{equation}
	Here the trace is taken over the qubit degrees of freedom, so what is left is the
	evolution only in the subspace of the environment. Note that such evolutions
	also lead to pure dephasing of the qubit, it is only that this process cannot
	be witnessed by any measurements on the environment.
	The other encompasses all types of evolutions which do involve evolution of the 
	environment, 
	\begin{equation}
	\Tr_Q\hat{\sigma}(t)=\hat{R}_{00}(t)=\hat{R}_{11}(t)=R(t)\neq \hat{R}(0).
	\end{equation}
	The density matrix of the environment conditional on the qubit being in state $|1\rangle$
	is defined as $\hat{R}_{11}(t)=\hat{w}_1(t)\hat{R}(0)\hat{w}_1^{\dagger}$
	in analogy to $\hat{R}_{00}(t)$.
	
	An evolution of the first category can never lead to echoed entanglement,
	since if $\hat{w}_0(t)\hat{R}(0)\hat{w}_0^{\dagger}=\hat{w}_1(t)\hat{R}(0)\hat{w}_1^{\dagger}
	=\hat{R}(0)$, we have
	\begin{eqnarray}
	\nonumber
	\hat{R}(0)&=&\hat{w}_1(t)\hat{R}(0)\hat{w}_1^{\dagger}
	=\hat{w}_1(t)\hat{w}_0(t)\hat{R}(0)\hat{w}_0^{\dagger}\hat{w}_1^{\dagger},\\
	\nonumber
	\hat{R}(0)&=&\hat{w}_0(t)\hat{R}(0)\hat{w}_0^{\dagger}
	=\hat{w}_0(t)\hat{w}_1(t)\hat{R}(0)\hat{w}_1^{\dagger}\hat{w}_0^{\dagger},
	\end{eqnarray}
	so the  separability criterion for the echoed state (\ref{cond_echo})
	is obviously fulfilled at all times without any additional assumption. Even 
	isolated instances of time, which would lead to entanglement in the echoed state
	for a separable
	pre-pulse state are impossible.
	
	In the other situation, we know that such instances of time exist, due to the
	examples above. To check if there exist time intervals in the pre-pulse evolution
	for which the echo generates entanglement, let us study a time interval $t\in[\tau_1,\tau_2]$ such that for any time $t$ within this interval we have
	$\hat{R}_{00}(t)=\hat{R}_{11}(t)$ (which guarantees pre-pulse separability).
	For there to be entanglement in the echoed state we need
	$\hat{w}_1(t)\hat{R}_{00}(t)\hat{w}_1^{\dagger}\neq\hat{w}_0(t)\hat{R}_{11}(t)\hat{w}_0^{\dagger}$,
	but because of the pre-pulse separability we can exchange the conditional environmental
	states and get $\hat{w}_1(t)\hat{R}_{11}(t)\hat{w}_1^{\dagger}\neq\hat{w}_0(t)\hat{R}_{00}(t)\hat{w}_0^{\dagger}$,
	or equivalently 
	\begin{equation}
	\hat{R}_{00}(2t)\neq \hat{R}_{11}(2t).
	\end{equation}
	Hence, for there to exist time-intervals for which the echo protocol leads to entanglement generation,
	the qubit-environment evolution without the echo procedure would have to fulfill
	a very specific requirement. Namely there would have to exist time intervals
	in which the evolution is separable, followed by time intervals in which QEE is generated.
	In other words, sudden birth of entanglement \cite{ficek08,mazzola09} would have to be possible in the system.
	
	The results of Ref.~\cite{Roszak_PRA18} show that for pure dephasing evolutions
	such as studied here,
	separability is equivalent to the lack of quantum discord \cite{ollivier01,henderson01,modi12} with respect to the environment.
	This means that the set of separable states has zero volume, and therefore sudden
	death of entanglement (which is a consequence of the geometry of separable states
	\cite{bengtsson06}) will not occur. Hence, also the transformation of separable
	evolutions to entangling ones via the quantum echo, when the evolution remains separable
	for finite or infinite time-intervals is not possible, and such occurrences are limited
	to isolated instances in time.
	
	\section{Echo signal as qubit-entanglement environment witness} \label{sec:witness}
	In the previous sections we have given examples showing that in general there is no correlation between the effectiveness of the echo protocol (measured by its capability to lead to coherence revival at time $2\tau$) and the generation of QEE. While this conclusion stands, as it is simply a manifestation of the fact that for an environment in a mixed state the correlation between amount of QEE and the strength of dephasing is rather weak, let us finish here with a more ``positive'' result for a specific case. 
	
Let us use the separability condition for the pre-pulse evolution of the qubit-environment system lasting for time $\tau$ in the form given by eq.~(\ref{wrw}).
Let us then focus on a qubit that couples to the environment in ``biased'' way \cite{Paz_PRA17,Kwiatkowski_PRB20}, so that $\hat{V}_0 \! =\ 0$ and only $\hat{V_1} \! =\! \lambda\hat{V}$ is nontrivial. 
This means that $\hat{R}_{00}(\tau) = \hat{R}(0)$, and QEE is generated if and only if $\hat{R}_{11}(\tau) \! \neq \! \hat{R}(0)$. A necessary condition for the latter is $[\hat{H}_1,\hat{R}(0)] \! \neq \! 0$. It is also a sufficient condition for QEE to appear at all $\tau$ but a subset of isolated points. This follows from an argument about impossibility of sudden death or birth of QEE from the previous Section: for $[\hat{H}_1,\hat{R}(0)] \! \neq \! 0$, QEE appears at the beginning of the evolution, and it cannot then vanish and stay zero for a finite stretch of time. 

We focus now on system in which the initial state of the environment is stationary
with respect to the free Hamiltonian of the environment, $[\hat{R}(0),\hat{H}_{E}] \! =\! 0$. The ``if and only if'' (with exception of isolated points in time) condition for nonzero QEE is then $[\hat{V}_1 , \hat{R}(0)]\! \neq \! 0$. A simple calculation of the commutator in expression for imaginary contribution to dephasing, eq.~(\ref{eq:Phit}), shows that the function $\Phi(2\tau)$ vanishes if the commutator of $\hat{V_1}$ and $\hat{R}(0)$ is zero. This leads to the following statement: if the environment is such a state, and the qubit's coupling is biased, the appearance of nonzero $\Phi(t)$ contribution to echo signal means that qubit and environment were entangled during the evolution (with possible exception of isolated points in time). This means that if the qubit is initialized with its Bloch vector in some direction (say $x$), then at echo time $2\tau$ the length of this vector is not only going to be diminished due to nonzero $\chi(2\tau)$, but due to nonzero $\Phi(t)$ the direction of the final vector is going to be rotated with respect to the original one. Under all the listed conditions, the appearance of such an environment-induced rotation of the echoed state of the qubit is equivalent to entangling nature of the evolution of the composite qubit-environment system.

	\section{Conclusion \label{sec5}}
	
	We have studied the spin echo performed on a qubit that interacts with an environment
	due to a type of Hamiltonian which leads to pure dephasing of the qubit. Our intent was
	to  find a relation between the performance of the echo procedure to reduce 
	decoherence, and the entanglement which can be generated between the qubit and
	its environment. Quite surprisingly, we have found that the effectiveness of the echo
	and entangement generation are two distinct issues. The perfect echo for which
	full coherence is restored can occur both in case of entangling and separable
	evolutions. 
	
	We have further analyzed the situation when the echo is not perfect, and found
	that it is possible for a qubit-environment state to be separable prior to the application of the local operation on the qubit (the $\pi$ pulse)
	while the final echoed state is entangled. It turns out that although such 
	a possibility does exist, it is limited to isolated instances of time.
	The important consequence here is that although the spin echo can result
	in the generation of entanglement from a point of time when there is no pre-pulse
	entanglement, this is a special case in an evolution which leads to entanglement
	generation on average. It cannot result in the change of the nature of evolution
	from nonentangling to entangling, so it cannot lead to a robust creation of 
	quantum correlations.
	
	Finally, we have shown that there is at least one case in which one can use the echo signal as a witness of the entangling charater of the evolution of a qubit and its environment. When the environment is initially in a stationary state with respect to the free Hamiltonian of the environment, and only one of two levels of the qubit is coupled to the environment (as happens for qubits for which only one of their levels has a finite dipole moment, e.g.~excitonic qubits \cite{borri01,vagov03,vagov04,roszak06b} or spin qubits based on $m = 0$ and $m=1$ levels of spin $S=1$ system, such as nitrogen-vacancy center \cite{Zhao_PRB12,Kwiatkowski_PRB18}). The appearance of phase shift of coherence \cite{Paz_PRA17,Kwiatkowski_PRB20} proves then the entangling nature of the evolution.

	\section{Acknowledgments}
	\L.~C. would like to thank Piotrek Sza\'nkowski for stimulating discussions. This work was funded from the Reseach Projects No.~UMO-2012/07/B/ST3/03616 and UMO-2015/19/B/ST3/03152 financed by the National Science Centre  of Poland (NCN).


\begin{thebibliography}{41}%
	\makeatletter
	\providecommand \@ifxundefined [1]{%
		\@ifx{#1\undefined}
	}%
	\providecommand \@ifnum [1]{%
		\ifnum #1\expandafter \@firstoftwo
		\else \expandafter \@secondoftwo
		\fi
	}%
	\providecommand \@ifx [1]{%
		\ifx #1\expandafter \@firstoftwo
		\else \expandafter \@secondoftwo
		\fi
	}%
	\providecommand \natexlab [1]{#1}%
	\providecommand \enquote  [1]{``#1''}%
	\providecommand \bibnamefont  [1]{#1}%
	\providecommand \bibfnamefont [1]{#1}%
	\providecommand \citenamefont [1]{#1}%
	\providecommand \href@noop [0]{\@secondoftwo}%
	\providecommand \href [0]{\begingroup \@sanitize@url \@href}%
	\providecommand \@href[1]{\@@startlink{#1}\@@href}%
	\providecommand \@@href[1]{\endgroup#1\@@endlink}%
	\providecommand \@sanitize@url [0]{\catcode `\\12\catcode `\$12\catcode
		`\&12\catcode `\#12\catcode `\^12\catcode `\_12\catcode `\%12\relax}%
	\providecommand \@@startlink[1]{}%
	\providecommand \@@endlink[0]{}%
	\providecommand \url  [0]{\begingroup\@sanitize@url \@url }%
	\providecommand \@url [1]{\endgroup\@href {#1}{\urlprefix }}%
	\providecommand \urlprefix  [0]{URL }%
	\providecommand \Eprint [0]{\href }%
	\providecommand \doibase [0]{http://dx.doi.org/}%
	\providecommand \selectlanguage [0]{\@gobble}%
	\providecommand \bibinfo  [0]{\@secondoftwo}%
	\providecommand \bibfield  [0]{\@secondoftwo}%
	\providecommand \translation [1]{[#1]}%
	\providecommand \BibitemOpen [0]{}%
	\providecommand \bibitemStop [0]{}%
	\providecommand \bibitemNoStop [0]{.\EOS\space}%
	\providecommand \EOS [0]{\spacefactor3000\relax}%
	\providecommand \BibitemShut  [1]{\csname bibitem#1\endcsname}%
	\let\auto@bib@innerbib\@empty
	\bibitem [{\citenamefont {Schlosshauer}(2007)}]{Schlosshauer_book}%
	\BibitemOpen
	\bibfield  {author} {\bibinfo {author} {\bibfnamefont {M.}~\bibnamefont
			{Schlosshauer}},\ }\href@noop {} {\emph {\bibinfo {title} {Decoherence and
				the Quantum-to-Classical Transition}}}\ (\bibinfo  {publisher} {Springer},\
	\bibinfo {address} {Berlin/Heidelberg},\ \bibinfo {year} {2007})\BibitemShut
	{NoStop}%
	\bibitem [{\citenamefont {Hornberger}(2009)}]{Hornberger}%
	\BibitemOpen
	\bibfield  {author} {\bibinfo {author} {\bibfnamefont {K.}~\bibnamefont
			{Hornberger}},\ }\bibfield  {title} {\enquote {\bibinfo {title} {Introduction
				to decoherence theory},}\ }in\ \href {\doibase 10.1007/978-3-540-88169-8_5}
	{\emph {\bibinfo {booktitle} {Entanglement and Decoherence}}},\ \bibinfo
	{series} {Lecture Notes in Physics}, Vol.\ \bibinfo {volume} {768},\ \bibinfo
	{editor} {edited by\ \bibinfo {editor} {\bibfnamefont {Andreas}\ \bibnamefont
			{Buchleitner}}, \bibinfo {editor} {\bibfnamefont {Carlos}\ \bibnamefont
			{Viviescas}}, \ and\ \bibinfo {editor} {\bibfnamefont {Markus}\ \bibnamefont
			{Tiersch}}}\ (\bibinfo  {publisher} {Springer Berlin Heidelberg},\ \bibinfo
	{year} {2009})\ pp.\ \bibinfo {pages} {221--276}\BibitemShut {NoStop}%
	\bibitem [{\citenamefont {{\.Z}urek}(2003)}]{Zurek_RMP03}%
	\BibitemOpen
	\bibfield  {author} {\bibinfo {author} {\bibfnamefont {Wojciech~Hubert}\
			\bibnamefont {{\.Z}urek}},\ }\bibfield  {title} {\enquote {\bibinfo {title}
			{Decoherence, einselection, and the quantum origins of the classical},}\
	}\href {\doibase 10.1103/RevModPhys.75.715} {\bibfield  {journal} {\bibinfo
			{journal} {Rev.\ Mod.\ Phys.}\ }\textbf {\bibinfo {volume} {75}},\ \bibinfo
		{pages} {715} (\bibinfo {year} {2003})}\BibitemShut {NoStop}%
	\bibitem [{\citenamefont {K{\"u}bler}\ and\ \citenamefont
		{Zeh}(1973)}]{Kuebler_AP73}%
	\BibitemOpen
	\bibfield  {author} {\bibinfo {author} {\bibfnamefont {O.}~\bibnamefont
			{K{\"u}bler}}\ and\ \bibinfo {author} {\bibfnamefont {H.~D.}\ \bibnamefont
			{Zeh}},\ }\bibfield  {title} {\enquote {\bibinfo {title} {Dynamics of quantum
				correlations},}\ }\href {\doibase 10.1016/0003-4916(73)90040-7} {\bibfield
		{journal} {\bibinfo  {journal} {Ann.~Phys.}\ }\textbf {\bibinfo {volume}
			{76}},\ \bibinfo {pages} {405} (\bibinfo {year} {1973})}\BibitemShut
	{NoStop}%
	\bibitem [{\citenamefont {Eisert}\ and\ \citenamefont
		{Plenio}(2002)}]{Eisert_PRL02}%
	\BibitemOpen
	\bibfield  {author} {\bibinfo {author} {\bibfnamefont {Jens}\ \bibnamefont
			{Eisert}}\ and\ \bibinfo {author} {\bibfnamefont {Martin~B.}\ \bibnamefont
			{Plenio}},\ }\bibfield  {title} {\enquote {\bibinfo {title} {Quantum and
				classical correlations in quantum brownian motion},}\ }\href {\doibase
		10.1103/PhysRevLett.89.137902} {\bibfield  {journal} {\bibinfo  {journal}
			{Phys. Rev. Lett.}\ }\textbf {\bibinfo {volume} {89}},\ \bibinfo {pages}
		{137902} (\bibinfo {year} {2002})}\BibitemShut {NoStop}%
	\bibitem [{\citenamefont {Hilt}\ and\ \citenamefont {Lutz}(2009)}]{Hilt_PRA09}%
	\BibitemOpen
	\bibfield  {author} {\bibinfo {author} {\bibfnamefont {Stefanie}\
			\bibnamefont {Hilt}}\ and\ \bibinfo {author} {\bibfnamefont {Eric}\
			\bibnamefont {Lutz}},\ }\bibfield  {title} {\enquote {\bibinfo {title}
			{System-bath entanglement in quantum thermodynamics},}\ }\href {\doibase
		10.1103/PhysRevA.79.010101} {\bibfield  {journal} {\bibinfo  {journal} {Phys.
				Rev. A}\ }\textbf {\bibinfo {volume} {79}},\ \bibinfo {pages} {010101}
		(\bibinfo {year} {2009})}\BibitemShut {NoStop}%
	\bibitem [{\citenamefont {Maziero}\ \emph {et~al.}(2010)\citenamefont
		{Maziero}, \citenamefont {Werlang}, \citenamefont {Fanchini}, \citenamefont
		{C\'eleri},\ and\ \citenamefont {Serra}}]{Maziero_PRA10}%
	\BibitemOpen
	\bibfield  {author} {\bibinfo {author} {\bibfnamefont {J.}~\bibnamefont
			{Maziero}}, \bibinfo {author} {\bibfnamefont {T.}~\bibnamefont {Werlang}},
		\bibinfo {author} {\bibfnamefont {F.~F.}\ \bibnamefont {Fanchini}}, \bibinfo
		{author} {\bibfnamefont {L.~C.}\ \bibnamefont {C\'eleri}}, \ and\ \bibinfo
		{author} {\bibfnamefont {R.~M.}\ \bibnamefont {Serra}},\ }\bibfield  {title}
	{\enquote {\bibinfo {title} {System-reservoir dynamics of quantum and
				classical correlations},}\ }\href {\doibase 10.1103/PhysRevA.81.022116}
	{\bibfield  {journal} {\bibinfo  {journal} {Phys. Rev. A}\ }\textbf {\bibinfo
			{volume} {81}},\ \bibinfo {pages} {022116} (\bibinfo {year}
		{2010})}\BibitemShut {NoStop}%
	\bibitem [{\citenamefont {Pernice}\ and\ \citenamefont
		{Strunz}(2011)}]{Pernice_PRA11}%
	\BibitemOpen
	\bibfield  {author} {\bibinfo {author} {\bibfnamefont {A.}~\bibnamefont
			{Pernice}}\ and\ \bibinfo {author} {\bibfnamefont {Walter~T.}\ \bibnamefont
			{Strunz}},\ }\bibfield  {title} {\enquote {\bibinfo {title} {Decoherence and
				the nature of system-environment correlations},}\ }\href {\doibase
		10.1103/PhysRevA.84.062121} {\bibfield  {journal} {\bibinfo  {journal} {Phys.
				Rev. A}\ }\textbf {\bibinfo {volume} {84}},\ \bibinfo {pages} {062121}
		(\bibinfo {year} {2011})}\BibitemShut {NoStop}%
	\bibitem [{\citenamefont {Roszak}\ and\ \citenamefont
		{Cywi{\'n}ski}(2015)}]{Roszak_PRA15}%
	\BibitemOpen
	\bibfield  {author} {\bibinfo {author} {\bibfnamefont {Katarzyna}\
			\bibnamefont {Roszak}}\ and\ \bibinfo {author} {\bibfnamefont {{\L}ukasz}\
			\bibnamefont {Cywi{\'n}ski}},\ }\bibfield  {title} {\enquote {\bibinfo
			{title} {Characterization and measurement of qubit-environment-entanglement
				generation during pure dephasing},}\ }\href {\doibase
		10.1103/PhysRevA.92.032310} {\bibfield  {journal} {\bibinfo  {journal} {Phys.
				Rev. A}\ }\textbf {\bibinfo {volume} {92}},\ \bibinfo {pages} {032310}
		(\bibinfo {year} {2015})}\BibitemShut {NoStop}%
	\bibitem [{\citenamefont {Roszak}(2018)}]{Roszak_qudit_PRA18}%
	\BibitemOpen
	\bibfield  {author} {\bibinfo {author} {\bibfnamefont {Katarzyna}\
			\bibnamefont {Roszak}},\ }\bibfield  {title} {\enquote {\bibinfo {title}
			{Criteria for system-environment entanglement generation for systems of any
				size in pure-dephasing evolutions},}\ }\href {\doibase
		10.1103/PhysRevA.98.052344} {\bibfield  {journal} {\bibinfo  {journal} {Phys.
				Rev. A}\ }\textbf {\bibinfo {volume} {98}},\ \bibinfo {pages} {052344}
		(\bibinfo {year} {2018})}\BibitemShut {NoStop}%
	\bibitem [{\citenamefont {Roszak}\ and\ \citenamefont
		{Cywi\'{n}ski}(2018)}]{Roszak_PRA18}%
	\BibitemOpen
	\bibfield  {author} {\bibinfo {author} {\bibfnamefont {Katarzyna}\
			\bibnamefont {Roszak}}\ and\ \bibinfo {author} {\bibfnamefont {\L{}ukasz}\
			\bibnamefont {Cywi\'{n}ski}},\ }\bibfield  {title} {\enquote {\bibinfo
			{title} {Equivalence of qubit-environment entanglement and discord generation
				via pure dephasing interactions and the resulting consequences},}\ }\href
	{\doibase 10.1103/PhysRevA.97.012306} {\bibfield  {journal} {\bibinfo
			{journal} {Phys. Rev. A}\ }\textbf {\bibinfo {volume} {97}},\ \bibinfo
		{pages} {012306} (\bibinfo {year} {2018})}\BibitemShut {NoStop}%
	\bibitem [{\citenamefont {Roszak}\ \emph {et~al.}(2019)\citenamefont {Roszak},
		\citenamefont {Kwiatkowski},\ and\ \citenamefont
		{Cywi\'{n}ski}}]{Roszak_PRA19}%
	\BibitemOpen
	\bibfield  {author} {\bibinfo {author} {\bibfnamefont {Katarzyna}\
			\bibnamefont {Roszak}}, \bibinfo {author} {\bibfnamefont {Damian}\
			\bibnamefont {Kwiatkowski}}, \ and\ \bibinfo {author} {\bibfnamefont
			{\L{}ukasz}\ \bibnamefont {Cywi\'{n}ski}},\ }\bibfield  {title} {\enquote
		{\bibinfo {title} {How to detect qubit-environment entanglement generated
				during qubit dephasing},}\ }\href {\doibase 10.1103/PhysRevA.100.022318}
	{\bibfield  {journal} {\bibinfo  {journal} {Phys. Rev. A}\ }\textbf {\bibinfo
			{volume} {100}},\ \bibinfo {pages} {022318} (\bibinfo {year}
		{2019})}\BibitemShut {NoStop}%
	\bibitem [{\citenamefont {Sza{\'n}kowski}\ and\ \citenamefont
		{Cywi{\'n}ski}(2020)}]{Szankowski_arXiv20}%
	\BibitemOpen
	\bibfield  {author} {\bibinfo {author} {\bibfnamefont {Piotr}\ \bibnamefont
			{Sza{\'n}kowski}}\ and\ \bibinfo {author} {\bibfnamefont {{\L}ukasz}\
			\bibnamefont {Cywi{\'n}ski}},\ }\bibfield  {title} {\enquote {\bibinfo
			{title} {Noise representations of open system dynamics},}\ }\href
	{http://arxiv.org/abs/2003.09688} {\bibfield  {journal} {\bibinfo  {journal}
			{arXiv:2003.09688}\ } (\bibinfo {year} {2020})}\BibitemShut {NoStop}%
	\bibitem [{\citenamefont {Hahn}(1950)}]{Hahn_PR50}%
	\BibitemOpen
	\bibfield  {author} {\bibinfo {author} {\bibfnamefont {E.~L.}\ \bibnamefont
			{Hahn}},\ }\href {\doibase 10.1103/PhysRev.80.580} {\bibfield  {journal}
		{\bibinfo  {journal} {Phys. Rev.}\ }\textbf {\bibinfo {volume} {80}},\
		\bibinfo {pages} {580} (\bibinfo {year} {1950})}\BibitemShut {NoStop}%
	\bibitem [{\citenamefont {Abragam}(1983)}]{Abragam}%
	\BibitemOpen
	\bibfield  {author} {\bibinfo {author} {\bibfnamefont {A.}~\bibnamefont
			{Abragam}},\ }\href@noop {} {\emph {\bibinfo {title} {The Principles of
				Nuclear Magnetism}}}\ (\bibinfo  {publisher} {Oxford University Press},\
	\bibinfo {address} {New York},\ \bibinfo {year} {1983})\BibitemShut {NoStop}%
	\bibitem [{\citenamefont {Vandersypen}\ and\ \citenamefont
		{Chuang}(2005)}]{Vandersypen_RMP05}%
	\BibitemOpen
	\bibfield  {author} {\bibinfo {author} {\bibfnamefont {L.~M.~K.}\
			\bibnamefont {Vandersypen}}\ and\ \bibinfo {author} {\bibfnamefont {I.~L.}\
			\bibnamefont {Chuang}},\ }\bibfield  {title} {\enquote {\bibinfo {title} {Nmr
				techniques for quantum control and computation},}\ }\href {\doibase
		10.1103/RevModPhys.76.1037} {\bibfield  {journal} {\bibinfo  {journal} {Rev.
				Mod. Phys.}\ }\textbf {\bibinfo {volume} {76}},\ \bibinfo {pages}
		{1037--1069} (\bibinfo {year} {2005})}\BibitemShut {NoStop}%
	\bibitem [{\citenamefont {de~Sousa}(2009)}]{deSousa_TAP09}%
	\BibitemOpen
	\bibfield  {author} {\bibinfo {author} {\bibfnamefont {Rogerio}\ \bibnamefont
			{de~Sousa}},\ }\bibfield  {title} {\enquote {\bibinfo {title} {Electron spin
				as a spectrometer of nuclear-spin noise and other fluctuations},}\ }\href
	{\doibase 10.1007/978-3-540-79365-610} {\bibfield  {journal} {\bibinfo
			{journal} {Top. Appl. Phys.}\ }\textbf {\bibinfo {volume} {115}},\ \bibinfo
		{pages} {183} (\bibinfo {year} {2009})}\BibitemShut {NoStop}%
	\bibitem [{\citenamefont {Sza\'nkowski}\ \emph {et~al.}(2017)\citenamefont
		{Sza\'nkowski}, \citenamefont {Ramon}, \citenamefont {Krzywda}, \citenamefont
		{Kwiatkowski},\ and\ \citenamefont {Cywi\'nski}}]{Szankowski_JPCM17}%
	\BibitemOpen
	\bibfield  {author} {\bibinfo {author} {\bibfnamefont {P.}~\bibnamefont
			{Sza\'nkowski}}, \bibinfo {author} {\bibfnamefont {G.}~\bibnamefont {Ramon}},
		\bibinfo {author} {\bibfnamefont {J.}~\bibnamefont {Krzywda}}, \bibinfo
		{author} {\bibfnamefont {D.}~\bibnamefont {Kwiatkowski}}, \ and\ \bibinfo
		{author} {\bibfnamefont {{\L}.}~\bibnamefont {Cywi\'nski}},\ }\bibfield
	{title} {\enquote {\bibinfo {title} {Environmental noise spectroscopy with
				qubits subjected to dynamical decoupling},}\ }\href {\doibase
		10.1088/1361-648X/aa7648} {\bibfield  {journal} {\bibinfo  {journal} {J.
				Phys.:Condens. Matter}\ }\textbf {\bibinfo {volume} {29}},\ \bibinfo {pages}
		{333001} (\bibinfo {year} {2017})}\BibitemShut {NoStop}%
	\bibitem [{\citenamefont {Chen}\ \emph {et~al.}(2018)\citenamefont {Chen},
		\citenamefont {Gneiting}, \citenamefont {Lo}, \citenamefont {Chen},\ and\
		\citenamefont {Nori}}]{chen18}%
	\BibitemOpen
	\bibfield  {author} {\bibinfo {author} {\bibfnamefont {Hong-Bin}\
			\bibnamefont {Chen}}, \bibinfo {author} {\bibfnamefont {Clemens}\
			\bibnamefont {Gneiting}}, \bibinfo {author} {\bibfnamefont {Ping-Yuan}\
			\bibnamefont {Lo}}, \bibinfo {author} {\bibfnamefont {Yueh-Nan}\ \bibnamefont
			{Chen}}, \ and\ \bibinfo {author} {\bibfnamefont {Franco}\ \bibnamefont
			{Nori}},\ }\bibfield  {title} {\enquote {\bibinfo {title} {Simulating open
				quantum systems with hamiltonian ensembles and the nonclassicality of the
				dynamics},}\ }\href {\doibase 10.1103/PhysRevLett.120.030403} {\bibfield
		{journal} {\bibinfo  {journal} {Phys. Rev. Lett.}\ }\textbf {\bibinfo
			{volume} {120}},\ \bibinfo {pages} {030403} (\bibinfo {year}
		{2018})}\BibitemShut {NoStop}%
	\bibitem [{\citenamefont {Chen}\ \emph {et~al.}(2019)\citenamefont {Chen},
		\citenamefont {Lo}, \citenamefont {Gneiting}, \citenamefont {Bae},
		\citenamefont {Chen},\ and\ \citenamefont {Nori}}]{chen19}%
	\BibitemOpen
	\bibfield  {author} {\bibinfo {author} {\bibfnamefont {Hong-Bin}\
			\bibnamefont {Chen}}, \bibinfo {author} {\bibfnamefont {Ping-Yuan}\
			\bibnamefont {Lo}}, \bibinfo {author} {\bibfnamefont {Clemens}\ \bibnamefont
			{Gneiting}}, \bibinfo {author} {\bibfnamefont {Joonwoo}\ \bibnamefont {Bae}},
		\bibinfo {author} {\bibfnamefont {Yueh-Nan}\ \bibnamefont {Chen}}, \ and\
		\bibinfo {author} {\bibfnamefont {Franco}\ \bibnamefont {Nori}},\ }\bibfield
	{title} {\enquote {\bibinfo {title} {Quantifying the nonclassicality of pure
				dephasing},}\ }\href {\doibase 10.1038/s41467-019-11502-4} {\bibfield
		{journal} {\bibinfo  {journal} {Nature Comm.}\ }\textbf {\bibinfo {volume}
			{10}},\ \bibinfo {pages} {3794} (\bibinfo {year} {2019})}\BibitemShut
	{NoStop}%
	\bibitem [{\citenamefont {Peres}(1996)}]{Peres_PRL96}%
	\BibitemOpen
	\bibfield  {author} {\bibinfo {author} {\bibfnamefont {Asher}\ \bibnamefont
			{Peres}},\ }\bibfield  {title} {\enquote {\bibinfo {title} {Separability
				criterion for density matrices},}\ }\href {\doibase
		10.1103/PhysRevLett.77.1413} {\bibfield  {journal} {\bibinfo  {journal}
			{Phys. Rev. Lett.}\ }\textbf {\bibinfo {volume} {77}},\ \bibinfo {pages}
		{1413--1415} (\bibinfo {year} {1996})}\BibitemShut {NoStop}%
	\bibitem [{\citenamefont {Horodecki}\ \emph {et~al.}(1996)\citenamefont
		{Horodecki}, \citenamefont {Horodecki},\ and\ \citenamefont
		{Horodecki}}]{Horodecki_PLA96}%
	\BibitemOpen
	\bibfield  {author} {\bibinfo {author} {\bibfnamefont {Micha{\l}}\
			\bibnamefont {Horodecki}}, \bibinfo {author} {\bibfnamefont {Pawe{\l}}\
			\bibnamefont {Horodecki}}, \ and\ \bibinfo {author} {\bibfnamefont {Ryszard}\
			\bibnamefont {Horodecki}},\ }\bibfield  {title} {\enquote {\bibinfo {title}
			{Separability of mixed states: necessary and sufficient conditions},}\ }\href
	{\doibase 10.1016/S0375-9601(96)00706-2} {\bibfield  {journal} {\bibinfo
			{journal} {Phys. Lett. A}\ }\textbf {\bibinfo {volume} {223}},\ \bibinfo
		{pages} {1--8} (\bibinfo {year} {1996})}\BibitemShut {NoStop}%
	\bibitem [{\citenamefont {Cywi{\'n}ski}(2011)}]{Cywinski_APPA11}%
	\BibitemOpen
	\bibfield  {author} {\bibinfo {author} {\bibfnamefont {{\L}ukasz}\
			\bibnamefont {Cywi{\'n}ski}},\ }\bibfield  {title} {\enquote {\bibinfo
			{title} {Dephasing of electron spin qubits due to their interaction with
				nuclei in quantum dots},}\ }\href {\doibase 10.12693/APhysPolA.119.576}
	{\bibfield  {journal} {\bibinfo  {journal} {Acta Phys.~Pol.~A}\ }\textbf
		{\bibinfo {volume} {119}},\ \bibinfo {pages} {576} (\bibinfo {year}
		{2011})}\BibitemShut {NoStop}%
	\bibitem [{\citenamefont {Chekhovich}\ \emph {et~al.}(2013)\citenamefont
		{Chekhovich}, \citenamefont {Makhonin}, \citenamefont {Tartakovskii},
		\citenamefont {Yacoby}, \citenamefont {Bluhm}, \citenamefont {Nowack},\ and\
		\citenamefont {Vandersypen}}]{Chekhovich_NM13}%
	\BibitemOpen
	\bibfield  {author} {\bibinfo {author} {\bibfnamefont {E.~A.}\ \bibnamefont
			{Chekhovich}}, \bibinfo {author} {\bibfnamefont {M.~N.}\ \bibnamefont
			{Makhonin}}, \bibinfo {author} {\bibfnamefont {A.~I.}\ \bibnamefont
			{Tartakovskii}}, \bibinfo {author} {\bibfnamefont {A.}~\bibnamefont
			{Yacoby}}, \bibinfo {author} {\bibfnamefont {H.}~\bibnamefont {Bluhm}},
		\bibinfo {author} {\bibfnamefont {K.~C.}\ \bibnamefont {Nowack}}, \ and\
		\bibinfo {author} {\bibfnamefont {L.~M.~K.}\ \bibnamefont {Vandersypen}},\
	}\bibfield  {title} {\enquote {\bibinfo {title} {Nuclear spin effects in
				semiconductor quantum dots},}\ }\href {\doibase 10.1038/nmat3652} {\bibfield
		{journal} {\bibinfo  {journal} {Nature Materials}\ }\textbf {\bibinfo
			{volume} {12}},\ \bibinfo {pages} {494} (\bibinfo {year} {2013})}\BibitemShut
	{NoStop}%
	\bibitem [{\citenamefont {Borri}\ \emph {et~al.}(2001)\citenamefont {Borri},
		\citenamefont {Langbein}, \citenamefont {Schneider}, \citenamefont {Woggon},
		\citenamefont {Sellin}, \citenamefont {Ouyang},\ and\ \citenamefont
		{Bimberg}}]{borri01}%
	\BibitemOpen
	\bibfield  {author} {\bibinfo {author} {\bibfnamefont {P.}~\bibnamefont
			{Borri}}, \bibinfo {author} {\bibfnamefont {W.}~\bibnamefont {Langbein}},
		\bibinfo {author} {\bibfnamefont {S.}~\bibnamefont {Schneider}}, \bibinfo
		{author} {\bibfnamefont {U.}~\bibnamefont {Woggon}}, \bibinfo {author}
		{\bibfnamefont {R.~L.}\ \bibnamefont {Sellin}}, \bibinfo {author}
		{\bibfnamefont {D.}~\bibnamefont {Ouyang}}, \ and\ \bibinfo {author}
		{\bibfnamefont {D.}~\bibnamefont {Bimberg}},\ }\bibfield  {title} {\enquote
		{\bibinfo {title} {Ultralong dephasing time in {InGaAs} quantum dots},}\
	}\href@noop {} {\bibfield  {journal} {\bibinfo  {journal} {Phys. Rev. Lett.}\
		}\textbf {\bibinfo {volume} {87}},\ \bibinfo {pages} {157401--1--4} (\bibinfo
		{year} {2001})}\BibitemShut {NoStop}%
	\bibitem [{\citenamefont {Vagov}\ \emph {et~al.}(2003)\citenamefont {Vagov},
		\citenamefont {Axt},\ and\ \citenamefont {Kuhn}}]{vagov03}%
	\BibitemOpen
	\bibfield  {author} {\bibinfo {author} {\bibfnamefont {A.}~\bibnamefont
			{Vagov}}, \bibinfo {author} {\bibfnamefont {V.~M.}\ \bibnamefont {Axt}}, \
		and\ \bibinfo {author} {\bibfnamefont {T.}~\bibnamefont {Kuhn}},\ }\bibfield
	{title} {\enquote {\bibinfo {title} {Impact of pure dephasing on the
				nonlinear optical response of single quantum dots and dot ensembles},}\
	}\href@noop {} {\bibfield  {journal} {\bibinfo  {journal} {Phys. Rev. B}\
		}\textbf {\bibinfo {volume} {67}},\ \bibinfo {pages} {115338} (\bibinfo
		{year} {2003})}\BibitemShut {NoStop}%
	\bibitem [{\citenamefont {Vagov}\ \emph {et~al.}(2004)\citenamefont {Vagov},
		\citenamefont {Axt}, \citenamefont {Kuhn}, \citenamefont {Langbein},
		\citenamefont {Borri},\ and\ \citenamefont {Woggon}}]{vagov04}%
	\BibitemOpen
	\bibfield  {author} {\bibinfo {author} {\bibfnamefont {A.}~\bibnamefont
			{Vagov}}, \bibinfo {author} {\bibfnamefont {V.~M.}\ \bibnamefont {Axt}},
		\bibinfo {author} {\bibfnamefont {T.}~\bibnamefont {Kuhn}}, \bibinfo {author}
		{\bibfnamefont {W.}~\bibnamefont {Langbein}}, \bibinfo {author}
		{\bibfnamefont {P.}~\bibnamefont {Borri}}, \ and\ \bibinfo {author}
		{\bibfnamefont {U.}~\bibnamefont {Woggon}},\ }\bibfield  {title} {\enquote
		{\bibinfo {title} {Nonmonotonous temperature dependence of the initial
				decoherence in quantum dots},}\ }\href@noop {} {\bibfield  {journal}
		{\bibinfo  {journal} {Phys. Rev. B}\ }\textbf {\bibinfo {volume} {70}},\
		\bibinfo {pages} {201305(R)--1--4} (\bibinfo {year} {2004})}\BibitemShut
	{NoStop}%
	\bibitem [{\citenamefont {Roszak}\ and\ \citenamefont
		{Machnikowski}(2006)}]{roszak06b}%
	\BibitemOpen
	\bibfield  {author} {\bibinfo {author} {\bibfnamefont {Katarzyna}\
			\bibnamefont {Roszak}}\ and\ \bibinfo {author} {\bibfnamefont {Pawe{\l}}\
			\bibnamefont {Machnikowski}},\ }\bibfield  {title} {\enquote {\bibinfo
			{title} {Complete disentanglement by partial pure dephasing},}\ }\href@noop
	{} {\bibfield  {journal} {\bibinfo  {journal} {Phys. Rev. A}\ }\textbf
		{\bibinfo {volume} {73}},\ \bibinfo {pages} {022313} (\bibinfo {year}
		{2006})}\BibitemShut {NoStop}%
	\bibitem [{\citenamefont {Zhao}\ \emph {et~al.}(2012)\citenamefont {Zhao},
		\citenamefont {Ho},\ and\ \citenamefont {Liu}}]{Zhao_PRB12}%
	\BibitemOpen
	\bibfield  {author} {\bibinfo {author} {\bibfnamefont {Nan}\ \bibnamefont
			{Zhao}}, \bibinfo {author} {\bibfnamefont {Sai-Wah}\ \bibnamefont {Ho}}, \
		and\ \bibinfo {author} {\bibfnamefont {Ren-Bao}\ \bibnamefont {Liu}},\
	}\bibfield  {title} {\enquote {\bibinfo {title} {Decoherence and dynamical
				decoupling control of nitrogen vacancy center electron spins in nuclear spin
				baths},}\ }\href {\doibase 10.1103/PhysRevB.85.115303} {\bibfield  {journal}
		{\bibinfo  {journal} {Phys. Rev. B}\ }\textbf {\bibinfo {volume} {85}},\
		\bibinfo {pages} {115303} (\bibinfo {year} {2012})}\BibitemShut {NoStop}%
	\bibitem [{\citenamefont {Kwiatkowski}\ and\ \citenamefont
		{Cywi\'{n}ski}(2018)}]{Kwiatkowski_PRB18}%
	\BibitemOpen
	\bibfield  {author} {\bibinfo {author} {\bibfnamefont {Damian}\ \bibnamefont
			{Kwiatkowski}}\ and\ \bibinfo {author} {\bibfnamefont {\L{}ukasz}\
			\bibnamefont {Cywi\'{n}ski}},\ }\bibfield  {title} {\enquote {\bibinfo
			{title} {Decoherence of two entangled spin qubits coupled to an interacting
				sparse nuclear spin bath: Application to nitrogen vacancy centers},}\ }\href
	{\doibase 10.1103/PhysRevB.98.155202} {\bibfield  {journal} {\bibinfo
			{journal} {Phys. Rev. B}\ }\textbf {\bibinfo {volume} {98}},\ \bibinfo
		{pages} {155202} (\bibinfo {year} {2018})}\BibitemShut {NoStop}%
	\bibitem [{\citenamefont {Paz-Silva}\ \emph {et~al.}(2017)\citenamefont
		{Paz-Silva}, \citenamefont {Norris},\ and\ \citenamefont
		{Viola}}]{Paz_PRA17}%
	\BibitemOpen
	\bibfield  {author} {\bibinfo {author} {\bibfnamefont {Gerardo~A.}\
			\bibnamefont {Paz-Silva}}, \bibinfo {author} {\bibfnamefont {Leigh~M.}\
			\bibnamefont {Norris}}, \ and\ \bibinfo {author} {\bibfnamefont {Lorenza}\
			\bibnamefont {Viola}},\ }\bibfield  {title} {\enquote {\bibinfo {title}
			{Multiqubit spectroscopy of gaussian quantum noise},}\ }\href {\doibase
		10.1103/PhysRevA.95.022121} {\bibfield  {journal} {\bibinfo  {journal} {Phys.
				Rev. A}\ }\textbf {\bibinfo {volume} {95}},\ \bibinfo {pages} {022121}
		(\bibinfo {year} {2017})}\BibitemShut {NoStop}%
	\bibitem [{\citenamefont {Kwiatkowski}\ \emph {et~al.}(2020)\citenamefont
		{Kwiatkowski}, \citenamefont {Sza{\'n}kowski},\ and\ \citenamefont
		{Cywi{\'n}ski}}]{Kwiatkowski_PRB20}%
	\BibitemOpen
	\bibfield  {author} {\bibinfo {author} {\bibfnamefont {D.}~\bibnamefont
			{Kwiatkowski}}, \bibinfo {author} {\bibfnamefont {P.}~\bibnamefont
			{Sza{\'n}kowski}}, \ and\ \bibinfo {author} {\bibfnamefont {\L.}\
			\bibnamefont {Cywi{\'n}ski}},\ }\bibfield  {title} {\enquote {\bibinfo
			{title} {Influence of nuclear spin polarization on the spin-echo signal of an
				nv-center qubit},}\ }\href {\doibase 10.1103/PhysRevB.101.155412} {\bibfield
		{journal} {\bibinfo  {journal} {Phys. Rev. B}\ }\textbf {\bibinfo {volume}
			{101}},\ \bibinfo {pages} {155412} (\bibinfo {year} {2020})}\BibitemShut
	{NoStop}%
	\bibitem [{\citenamefont {Negele}\ and\ \citenamefont {Orland}(1988)}]{Negele}%
	\BibitemOpen
	\bibfield  {author} {\bibinfo {author} {\bibfnamefont {John~W.}\ \bibnamefont
			{Negele}}\ and\ \bibinfo {author} {\bibfnamefont {Henri}\ \bibnamefont
			{Orland}},\ }\href@noop {} {\emph {\bibinfo {title} {Quantum Many-Particle
				Systems}}}\ (\bibinfo  {publisher} {Addison-Wesley},\ \bibinfo {address}
	{Redwood City, CA},\ \bibinfo {year} {1988})\BibitemShut {NoStop}%
	\bibitem [{\citenamefont {Bruus}\ and\ \citenamefont
		{Flensberg}(2004)}]{Bruus}%
	\BibitemOpen
	\bibfield  {author} {\bibinfo {author} {\bibfnamefont {H.}~\bibnamefont
			{Bruus}}\ and\ \bibinfo {author} {\bibfnamefont {K.}~\bibnamefont
			{Flensberg}},\ }\href@noop {} {\emph {\bibinfo {title} {Many-Body Quantum
				Field Theory in Condensed Matter Physics}}}\ (\bibinfo  {publisher} {Oxford
		University Press},\ \bibinfo {address} {Oxford},\ \bibinfo {year}
	{2004})\BibitemShut {NoStop}%
	\bibitem [{\citenamefont {Cywi{\'n}ski}\ \emph {et~al.}(2008)\citenamefont
		{Cywi{\'n}ski}, \citenamefont {Lutchyn}, \citenamefont {Nave},\ and\
		\citenamefont {{Das Sarma}}}]{Cywinski_PRB08}%
	\BibitemOpen
	\bibfield  {author} {\bibinfo {author} {\bibfnamefont {{\L}ukasz}\
			\bibnamefont {Cywi{\'n}ski}}, \bibinfo {author} {\bibfnamefont {Roman~M.}\
			\bibnamefont {Lutchyn}}, \bibinfo {author} {\bibfnamefont {Cody~P.}\
			\bibnamefont {Nave}}, \ and\ \bibinfo {author} {\bibfnamefont
			{S.}~\bibnamefont {{Das Sarma}}},\ }\bibfield  {title} {\enquote {\bibinfo
			{title} {How to enhance dephasing time in superconducting qubits},}\ }\href
	{\doibase 10.1103/PhysRevB.77.174509} {\bibfield  {journal} {\bibinfo
			{journal} {Phys.\ Rev.\ B}\ }\textbf {\bibinfo {volume} {77}},\ \bibinfo
		{pages} {174509} (\bibinfo {year} {2008})}\BibitemShut {NoStop}%
	\bibitem [{\citenamefont {Ficek}\ and\ \citenamefont
		{Tana\ifmmode~\acute{s}\else \'{s}\fi{}}(2008)}]{ficek08}%
	\BibitemOpen
	\bibfield  {author} {\bibinfo {author} {\bibfnamefont {Zbigniew}\
			\bibnamefont {Ficek}}\ and\ \bibinfo {author} {\bibfnamefont {Ryszard}\
			\bibnamefont {Tana\ifmmode~\acute{s}\else \'{s}\fi{}}},\ }\bibfield  {title}
	{\enquote {\bibinfo {title} {Delayed sudden birth of entanglement},}\ }\href
	{\doibase 10.1103/PhysRevA.77.054301} {\bibfield  {journal} {\bibinfo
			{journal} {Phys. Rev. A}\ }\textbf {\bibinfo {volume} {77}},\ \bibinfo
		{pages} {054301} (\bibinfo {year} {2008})}\BibitemShut {NoStop}%
	\bibitem [{\citenamefont {Mazzola}\ \emph {et~al.}(2009)\citenamefont
		{Mazzola}, \citenamefont {Maniscalco}, \citenamefont {Piilo}, \citenamefont
		{Suominen},\ and\ \citenamefont {Garraway}}]{mazzola09}%
	\BibitemOpen
	\bibfield  {author} {\bibinfo {author} {\bibfnamefont {L.}~\bibnamefont
			{Mazzola}}, \bibinfo {author} {\bibfnamefont {S.}~\bibnamefont {Maniscalco}},
		\bibinfo {author} {\bibfnamefont {J.}~\bibnamefont {Piilo}}, \bibinfo
		{author} {\bibfnamefont {K.-A.}\ \bibnamefont {Suominen}}, \ and\ \bibinfo
		{author} {\bibfnamefont {B.~M.}\ \bibnamefont {Garraway}},\ }\bibfield
	{title} {\enquote {\bibinfo {title} {Sudden death and sudden birth of
				entanglement in common structured reservoirs},}\ }\href {\doibase
		10.1103/PhysRevA.79.042302} {\bibfield  {journal} {\bibinfo  {journal} {Phys.
				Rev. A}\ }\textbf {\bibinfo {volume} {79}},\ \bibinfo {pages} {042302}
		(\bibinfo {year} {2009})}\BibitemShut {NoStop}%
	\bibitem [{\citenamefont {{Ollivier}}\ and\ \citenamefont
		{{Zurek}}(2002)}]{ollivier01}%
	\BibitemOpen
	\bibfield  {author} {\bibinfo {author} {\bibfnamefont {H.}~\bibnamefont
			{{Ollivier}}}\ and\ \bibinfo {author} {\bibfnamefont {W.~H.}\ \bibnamefont
			{{Zurek}}},\ }\bibfield  {title} {\enquote {\bibinfo {title} {{Quantum
					Discord: A Measure of the Quantumness of Correlations}},}\ }\href {\doibase
		10.1103/PhysRevLett.88.017901} {\bibfield  {journal} {\bibinfo  {journal}
			{Physical Review Letters}\ }\textbf {\bibinfo {volume} {88}},\ \bibinfo {eid}
		{017901} (\bibinfo {year} {2002})},\ \Eprint
	{http://arxiv.org/abs/quant-ph/0105072} {quant-ph/0105072} \BibitemShut
	{NoStop}%
	\bibitem [{\citenamefont {{Henderson}}\ and\ \citenamefont
		{{Vedral}}(2001)}]{henderson01}%
	\BibitemOpen
	\bibfield  {author} {\bibinfo {author} {\bibfnamefont {L.}~\bibnamefont
			{{Henderson}}}\ and\ \bibinfo {author} {\bibfnamefont {V.}~\bibnamefont
			{{Vedral}}},\ }\bibfield  {title} {\enquote {\bibinfo {title} {{Classical,
					quantum and total correlations}},}\ }\href {\doibase
		10.1088/0305-4470/34/35/315} {\bibfield  {journal} {\bibinfo  {journal}
			{Journal of Physics A Mathematical General}\ }\textbf {\bibinfo {volume}
			{34}},\ \bibinfo {pages} {6899--6905} (\bibinfo {year} {2001})},\ \Eprint
	{http://arxiv.org/abs/arXiv:quant-ph/0105028} {arXiv:quant-ph/0105028}
	\BibitemShut {NoStop}%
	\bibitem [{\citenamefont {{Modi}}\ \emph {et~al.}(2012)\citenamefont {{Modi}},
		\citenamefont {{Brodutch}}, \citenamefont {{Cable}}, \citenamefont
		{{Paterek}},\ and\ \citenamefont {{Vedral}}}]{modi12}%
	\BibitemOpen
	\bibfield  {author} {\bibinfo {author} {\bibfnamefont {K.}~\bibnamefont
			{{Modi}}}, \bibinfo {author} {\bibfnamefont {A.}~\bibnamefont {{Brodutch}}},
		\bibinfo {author} {\bibfnamefont {H.}~\bibnamefont {{Cable}}}, \bibinfo
		{author} {\bibfnamefont {T.}~\bibnamefont {{Paterek}}}, \ and\ \bibinfo
		{author} {\bibfnamefont {V.}~\bibnamefont {{Vedral}}},\ }\bibfield  {title}
	{\enquote {\bibinfo {title} {{The classical-quantum boundary for
					correlations: Discord and related measures}},}\ }\href {\doibase
		10.1103/RevModPhys.84.1655} {\bibfield  {journal} {\bibinfo  {journal}
			{Reviews of Modern Physics}\ }\textbf {\bibinfo {volume} {84}},\ \bibinfo
		{pages} {1655--1707} (\bibinfo {year} {2012})},\ \Eprint
	{http://arxiv.org/abs/1112.6238} {arXiv:1112.6238 [quant-ph]} \BibitemShut
	{NoStop}%
	\bibitem [{\citenamefont {Bengtsson}\ and\ \citenamefont
		{Zyczkowski}(2006)}]{bengtsson06}%
	\BibitemOpen
	\bibfield  {author} {\bibinfo {author} {\bibfnamefont {Ingemar}\ \bibnamefont
			{Bengtsson}}\ and\ \bibinfo {author} {\bibfnamefont {Karol}\ \bibnamefont
			{Zyczkowski}},\ }\href {\doibase 10.1017/CBO9780511535048} {\emph {\bibinfo
			{title} {Geometry of Quantum States: An Introduction to Quantum
				Entanglement}}}\ (\bibinfo  {publisher} {Cambridge University Press},\
	\bibinfo {year} {2006})\BibitemShut {NoStop}%
\end{thebibliography}

%

\end{document}